\documentclass[runningheads,a4paper]{llncs}

\usepackage[english]{babel}

\usepackage{booktabs}

\usepackage[%
rm={oldstyle=false,proportional=true},%
sf={oldstyle=false,proportional=true},%
tt={oldstyle=false,proportional=true,variable=true},%
qt=false%
]{cfr-lm}
\usepackage{graphicx}

\usepackage{paralist}
\usepackage{graphicx}
\usepackage{subcaption}

\usepackage{cite}

\usepackage[T1]{fontenc}

\usepackage[math]{blindtext}

\usepackage{csquotes}

\usepackage{microtype}

\usepackage{url}
\makeatletter
\g@addto@macro{\UrlBreaks}{\UrlOrds}
\makeatother

\usepackage[table,xcdraw]{xcolor}

\usepackage[
bookmarks=false,
breaklinks=true,
colorlinks=true,
linkcolor=black,
citecolor=black,
urlcolor=black,
pdfpagelayout=SinglePage,
pdfstartview=Fit
]{hyperref}
\usepackage[all]{hypcap}

\usepackage{todonotes}

\usepackage{xspace}
\newcommand{\eg}{{\em e.\,g.,\/}}
\newcommand{\ie}{{\em i.\,e.,\/}}
\newcommand{\etc}{{\em etc.\/}}

\DeclareFontFamily{U}{MnSymbolC}{}
\DeclareSymbolFont{MnSyC}{U}{MnSymbolC}{m}{n}
\DeclareFontShape{U}{MnSymbolC}{m}{n}{
    <-6>  MnSymbolC5
   <6-7>  MnSymbolC6
   <7-8>  MnSymbolC7
   <8-9>  MnSymbolC8
   <9-10> MnSymbolC9
  <10-12> MnSymbolC10
  <12->   MnSymbolC12%
}{}
\DeclareMathSymbol{\powerset}{\mathord}{MnSyC}{180}

\usepackage{soul}

\usepackage[acronym]{glossaries} 
\newacronym{ast}{AST}{Abstract Syntax Tree}
\newacronym{ct}{CT}{Combinatorial Testing}
\newacronym{czt}{CZT}{Community Z Tools}
\newacronym{dap}{DAP}{Debug Adapter Protocol}
\newacronym{dbgp}{DBGP}{Common DeBugGer Protocol}
\newacronym{gui}{GUI}{Graphical User Interface}
\newacronym{ide}{IDE}{Integrated Development Environment}
\newacronym{lsp}{LSP}{Language Server Protocol}
\newacronym{poc}{PoC}{Proof of Concept}
\newacronym{pog}{POG}{Proof Obligation Generation}
\newacronym{po}{PO}{Proof Obligation}
\newacronym{slsp}{SLSP}{Specification Language Server Protocol}
\newacronym{vdm}{VDM}{Vienna Development Method}
\newacronym{vscode}{VS Code}{Visual Studio Code}
\newacronym{CSV}{CSV}{Comma Separated Values}
\newacronym{ISQ}{ISQ}{International System of Quantities}

\usepackage{chngcntr}

\usepackage[capitalise,nameinlink]{cleveref}
\crefname{section}{Sect.}{Sect.}
\Crefname{section}{Section}{Sections}
\crefname{lstlisting}{Listing}{Listing}
\Crefname{lstlisting}{Listing}{Listing}

\usepackage{listings} 
\usepackage{times}    

\lstdefinelanguage{json}{
    basicstyle=\ttfamily\small, 
    numbers=left,
    stepnumber=1,
    numbersep=8pt,
    breaklines=true,
    frame=single,
    xleftmargin=.11\textwidth, 
    xrightmargin=.11\textwidth
}

\usepackage{vdmlisting}

\begin{document}
\pdfgentounicode=1

\title{Topologically sorting VDM-SL definitions for Isabelle/HOL translation}

\author{Leo Freitas\inst{1} \and Nick Battle\inst{2}
}
\authorrunning{ }

\institute{
School of Computing, Newcastle University, 
\email{leo.freitas@newcastle.ac.uk}
\and
Independent, \email{nick.battle@acm.org}
}
			
\maketitle
\setcounter{footnote}{0} 
\begin{abstract}
    There is an ecosystem of VDM libraries and extensions that includes a translation and proof environment for VDM in Isabelle. Translation works for a large subset of VDM-SL and further constructs are being added on demand. A key impediment for novice users is the fact Isabelle/HOL requires all definitions to be declared before they are used, where (mutually) recursive definitions \textbf{must} be defined in tandem. In this paper, we describe a solution to this problem, which will enable wider access to the translator plugin for novice users as well real models.   
\end{abstract}


\section{Introduction}\label{sec:intro}

The \gls{vdm} has has been widely used both in industrial contexts and academic ones covering several domains of the fields of Security~\cite{Kulik&20,Kulik&21a}, Fault-Tolerance~\cite{Nilsson&18}, Medical Devices~\cite{Macedo&08}, among others. We extend VDM specification support with a suite of tools and mathematical libraries~\footnote{\url{https://github.com/leouk/VDM_Toolkit/}}. 

VDM also has a \gls{vscode} IDE with multiple features, including a translation strategy to Isabelle/HOL~\cite{AdvancedVSCodePaper}. Nevertheless, the user needs to follow a strict and specific specification style, otherwise translation will either fail or be impossible. For example, every Isabelle/HOL theory \textbf{must} be written within a file of the same name, whereas VDM files might have no modules or multiple modules. Translation of such VDM specifications ought to impose the required Isabelle style. Currently, users have to be aware of such issues, or else translation will fail.   

There are various aspects of these issues, which are listed in various example files in the distribution, and explained in detail in~\cite{NimFull}. These VDM ``idioms'' are important not only to enable translation, but also to ensure proofs will be manageable and proof strategies will be easier to identify. 

Nevertheless, a major impediment for Isabelle translation is the fact VDM specifications \textbf{must} be ordered (\ie have declarations defined before they are used). This is rather unnatural to most VDM users, who tend to write specifications top-down, from larger/complex concepts to simpler/easier definitions. Also, it makes translation of legacy models hard because they were not written with such order requirement in mind. Manual rearrangement is not difficult, but it is laborious and error prone, hence creating a barrier to entry for the translator. 

Solving this specific order requirement is the key contribution of this paper. We present the solution following the style by Naur's N-Queens algorithm~\cite{NQueens}:~we provide a historical account to how various developments within the VDM tools eventually led to the possibility of module sorting. These are spread across various sections, including a historical overview of developments.  

\section{Background}~\label{sec:background}

VDM translation to Isabelle/HOL is performed by traversing its abstract syntax tree (AST) and issuing corresponding Isabelle/HOL definitions, for every part of the VDM concrete syntax that is compatible for translation. This process can be complex. For example, VDM union types are quite expressive and can allow some peculiar definitions, like this cross union selection example.
\begin{vdmsl}[frame=none,basicstyle=\ttfamily\scriptsize]
    types
        TUnion1 = int | nat
        inv u == (is_nat(u) => u > 0) and (is_int(u) => u < 0);

        TUnion2 = seq of nat | set of real;

    functions 
        f: TUnion2 * TUnion1 -> bool
        f(u1, u2) ==
            (is_int(u2) => 
                ((is_(u1, seq of nat) => u2 in set elems u1)
                  and
                 (is_(u1, set of real) => u2 in set u1))
            )
            and
            (is_nat(u2) => 
                ((is_(u1, seq of nat) => not u2 in set elems u1)
                  and
                 (is_(u1, set of real) => not u2 in set u1))
            );
\end{vdmsl}
VDM union types will create a maximal type set over the united types, where invariants are preserved. This can create quite complex (and confusing) selection and invariant checking processes. Of course, such specifications are rarely written. Yet, from a translation point of view, this illustrates some of the challenges around VDM's expressivity with respect to how they can be translated to Isabelle. 

Another rather complex definition space in VDM are those for values and patterns. For example, one could define some complex VDM patterns in values and types like:
\begin{vdmsl}[frame=none,basicstyle=\ttfamily\scriptsize]
    types
        T = set of nat inv {a,b,c} == a < b and b < c;
    values 
        [i,j]: seq of nat = [1,2];
        {k,m}: set of nat = {2,1};
\end{vdmsl}
The type definition states it is an ordered set of three explicit \verb'nat' values, whereas the value definitions bind each name to its corresponding value (\ie~\verb'i=1', \verb'j=2', where \verb'k' and \verb'm' will be a random choice within the set!). This poses various challenges for translation, given that Isabelle requires stricter bindings between names, types and values. Practically, such examples are rarely an impediment (or rarely written), yet have to be handled by the sorting algorithm as well. In particular, value definitions like those of \verb'[i,j]', effectively create two separate value definitions for both \verb'i' and \verb'j'.   

The translator, therefore, caters for the subset of VDM that is ``tamer'' (and mostly used). The translator has a VDM analysis tool (named \texttt{exu}\footnote{The name stems from the Yoruba divinity \textit{\`{E}\v{s}\`{u}} that was the gate keeper (and messenger) between mortals and deities.}), which checks for various such syntactic conondrums and limitations, telling users what to do and where problems are. Fixing such errors and warnings is important to ensure that translations are possible and without type errors for the user to solve on their own.  

\texttt{exu} also provides support to prepare users for translation (\eg~there are over \(50\) kinds of error and \(20\) warnings that \texttt{exu} can issue). For example, it checks that any function call within a function definition ought to include that function's precondition, if one exists. Yet, one key hindrance remains:~ordering of module definitions and their dependencies. This hindrance will be addressed in this paper and has recently been implemented in the latest version of the translator (see Section~\ref{sec:integration}). 


\section{Important Historical Developments}\label{sec:history}

Before getting to our solution, it is important to identify key (albeit techically unrelated) extensions to the main VDM tools. They were the stepping stones that eventually enabled \texttt{exu} to sort definitions for translation. We highlight the motivation for such extensions, as well as how they are implemented, in the main VDM tools below.

\subsection{Load time problems}

When working with large VDM models~\cite{emv2} (\eg~\(150\)+ modules, \(60\)+ KLOC), typecheck time (\eg~\(2-4\) minutes) can become a hindrance, and initialisation time (\eg~\(15\)+ minutes) can compromise further development. 

Such long initialization times occurred because the VDM runtime performs multiple passes through the AST in order to cater for forward references. The number of passes depends on the number of module inter-dependencies (\ie~\texttt{module A} imports \texttt{module B} and vice-versa), and the order of definitions within modules (\ie~using definitions before defining them), etc. 

To optimise type checking and initialization, users of large models would have to keep track of dependencies by carefully checking import chains and intra-module definitions. For example, users would have to have carefully crafted dependencies and keep their documentation explicitly. This was error prone and hard to maintain. 

\paragraph*{Verbose output.}
To address this, an extension was added to VDMJ~\cite{Battle09} that tracked top-level dependencies across modules. A simple and easy solution was for the typechecker to issue verbose output about every top-level definition's dependencies as it traversed the VDM AST, looking for cyclic dependencies. This verbose output enabled users to know how many passes had taken place, and whether cyclic dependencies existed and where. Users could then fix the cycles manually. This meant one could use the verbose output to minimise passes by lowering forward references, hence minimising load time.   

\paragraph*{Ordering, using dependency and free variable visitors.}
For identifying the optimal module ordering in VDM-SL, it was sufficient to consider each module's imports. This created a dependency graph and the objective was to process the ``most depended on'' modules first. In VDM++, classes do not have explicit imports, but rather can use any public definitions from other classes. So class ordering used the same visitor that the type checker uses to search for cyclic dependencies - that is, it searches for which other public definitions the current one depends on for initialization. The objective was to process the ``most depended on'' class files first. Lastly, within either a module or a class, the optimal definition ordering minimises the number of forward references. This required a different visitor, which looks for the free variables used by each definition. The cyclic dependency visitor and the free variable visitor were very similar, but the cyclic dependency case \emph{must not produce false positives}, \ie~claiming that there is a cycle when there is not. This means that undecidable cases, such as cycles that include conditional ``if'' statements, must be ignored. On the other hand, the free variable visitor is only concerned with ordering at analysis-time and can tolerate cycles, so it returns all free variables.

\paragraph*{Topological sort of module dependencies.}
With the information gathered about dependencies, it was possible to create a topological sort of module names, where the result would be a list of module names with the fewest passes possible. This effectively solved (in most cases) the dependency warnings from the verbose output, hence giving the end user the ordered list of module names to load. This could then be given to VDMJ at load time directly. This has enabled output of an acyclic directed graph view of module dependencies as a \texttt{dot} file.    

\subsection{Recursive cycles detection and removal.}

In general, a dependency graph can have cycles. This is not necessarily an error, in VDM. For example, it is perfectly reasonable to have mutually recursive functions or mutually dependent modules; but it is not legal to have mutually dependent value initializations (\eg~ A = B and B = A). So the VDMJ typechecker looks for initialization dependency cycles and treats these as errors; the same dependency search was used to support the optimal ordering of modules or classes.

But to produce an ordered sequence of nodes from a graph, it first had to be reduced to an acyclic graph (a DAG). This means that legitimate cycles had to be broken in such a way that the ordering of the resulting DAG was still optimal - in our case, has the fewest forward references. This problem is hard (it is known to be NP-Hard); it is complicated by the fact that definitions within a source file cannot (currently) be re-ordered by the typechecker\footnote{This could be a future enhancement to VDMJ.}, so the graph has fixed subgraphs (source file definitions) whose dependencies cannot be changed, and inter-subgraph dependencies (between files) that can be changed.

The solution adopted by VDMJ's module/class ordering was naive, simply removing the last link found that leads to a cycle, repeatedly, until no cycles exist. This was fast, and produced a vastly improved ordering in most cases (\ie~far better than guesswork); in the worst case, a suboptimal module order would still typecheck and initialize correctly, so this was not critical. However, in the case of Isabelle definition ordering, forward references are an error, so care would have to be taken with cycle removal and DAG startpoint nodes selection.

\section{Exu ordering}\label{sec:exu}

A topological sort of dependencies between modules resolved the long typecheck and initialization time issues. However, it did not optimize definitions within modules themselves. Moreover, the module dependency search only considered imports. It did not consider local (\eg~\texttt{let-in} definitions) and function namespaces, where type namespaces were not checked. It was also related to top-level imported definitions (\ie~types, values, functions, \etc) rather than their structure.    This required a different topological sort algorithm. The new algorithm had five stages: 
\begin{enumerate} 
    \item Collect all named definitions;
    \item Process non-function space dependenceies:
        \begin{itemize}
            \item visit all definition dependencies; 
            \item create any missing \verb'inv_T', for all declared types \verb'T';
            \item link type and definition spaces; 
        \end{itemize}
    \item Process function definition name-space dependencies:
        \begin{itemize}
            \item visit name-space dependencies;
            \item ignore recursive calls (those are  handled by VDMJ's cyclic dependency checks);
            \item link named dependencies;  
        \end{itemize}
    \item Topological sort:
        \begin{itemize}
            \item Checks whether topological sorting is needed;
            \item Kahn's algorithm DAG sorting of top-level names; 
         \end{itemize}
    \item Module reconstruction:
         \begin{itemize}
             \item Organise top-level names to separate type from function name-spaces;  
             \item Reorder module definitions;
             \item Optionally re-typechecks module; 
          \end{itemize}
\end{enumerate}  
The first stage effectively flattens the module top-level definitions into a single list of everything being defined within the module. For example, explicit functions might define pre/postconditions as well as measures, whereas types might define invariant, equality or order predicates. Each of these definitions are explicit function themselves, which have to be explicitly processed for (free-variable) forward dependenceies. 

The second stage searches for top-level definition dependencies, where it creates invariant function definitions for all types that do not have them explicitly declared. This is important to ensure that every type \verb'T' without explicit state invariants still depend on the function \verb'inv_T'. This effectively creates a function-space dependency between user-declared types, values and functions. Such synthetic constructs are structurally equivalent to what the typechecker would create had the user written something like \verb'inv_T = true' explicitly. This caters for the way both records and named typed invariant functions are constructed. For example, records create a function from the record name to boolean, whereas a named type \verb'S = RHS' creates a function from the \verb'RHS' type to boolean. Next, for declared values, there are no implicit functions to create, yet the named (type) dependency link is noted. This deeply processes values, which can have multiple local definitions, as described in the examples above in Section~\ref{sec:background}.    

The third stage searches for free-variable dependencies across all top-level definitions, taking into account local contexts, such as function parameters or locally bound variables (\eg~quantified variables, let-def constructs, \etc). This search also finds recursive, as well as mutually recursive, function calls. The former are ignored/removed, given they are treated differently by the translator. The latter must be resolved, otherwise the topological search stage will fail, given it requires a directed acyclic graph (DAG).    

The fourth stage is the actual topological sort algorithm used by VDMJ to sort module names described above. It is a variation of Kahn's DAG sort algorithm\footnote{\url{https://en.wikipedia.org/wiki/Topological_sorting\#Kahn's_algorithm}}. The DAG sort algorithm also looks for (and does not tolerate) cyclic dependencies within the given graph. Such cyclic dependencies have to be resolved earlier. In the case of definitions, cyclic dependencies can only occur between mutually recursive type of function definitions. Fortunately, VDMJ already has a mechanism for removing cyclic dependencies, which we apply here as well. This works well in practice. In the general case, however, this might fail, depending on the complexity of the mutual recursion(s) involved (see above).

The first three stages are considered as the module definition processing phase. They create all the necessary structures and implicit definitions needed for sorting. Nevertheless, before the sorting itself, it is important to provide two sources of information. First, the sort has to be made based on so-called graph start points:~nodes without incoming links. Second, sorting is only to be attempted if the module does contain use of definitions before their declaration. That is because attempting to sort an already sorted module can lead to an inconsistent result (\ie~a sorting order that does not prevent use before declaration). Thus, we ensure that Kahn's algorithm is only run when necessary. Whether it will be necessary is discovered based on a variation of VDMJ's verbose output warning for cyclic dependencies check.

Finally, for the fifth stage, if module sorting is not required, then the topological sorting is skipped and the module is not modified. Otherwise, if the sorting is required, then it is necessary to reorganise the topologically sorted list of names to take into account the two separate names spaces for types and functions at a module top-level definitions (prior to flattening definitions in the first stage). 

For example, consider the module \verb'M' below. It defines types and functions top-down, in use before declaration fashion, where some definitions do not have explicitly defined invariants. 
\begin{vdmsl}[frame=none,basicstyle=\ttfamily\scriptsize,numbers=left,stepnumber=1]
    module M 
    exports all
    definitions 
    types 
        --@doc uses types S and T before their declaration; implicit inv_Rec needed
        Rec:: s: S t:T;
        
        --@doc uses type T and function tail before declaring them.
        S = T inv s == head(s) > 0 and len tail(s) > 0;
        
        --@doc implicit inv_T needed
        T = seq1 of nat;

    functions
        tail: seq1 of nat -> seq of nat
        tail(s) == tl s;

        head: seq1 of nat -> nat
        head(s) == hd s;
    end M
\end{vdmsl}
When users run \verb'exu' (with the \verb'debug' flag set) on this module this is the tool's output. 
\begin{vdmsl}[frame=none,basicstyle=\ttfamily\scriptsize]
    Calling Exu VDM analyser...
    Calculating declaration dependencies for module `M`...
    Printed dependencies for module M.dot at ./M.dot
    M`T declared after Rec
    M`S declared after Rec
    M`T declared after S
    tail declared after S
    tail declared after inv_S
    Found 5 definition use before declaration. Topological sorted required.
    Original names : Rec, S, T, tail, head
    Start points   : Rec, S
    Sorted names   : tail, head, inv_T, inv_S, T, inv_Rec, S, Rec
    Organised names: tail, head, T, S, Rec
    Exu successfully sorted module M definitions
\end{vdmsl}
The original names are the user defined top-level names in declaration order, which are available after the first stage. The start points are those names at the top of the graph (\ie~with no incoming links), which are available after the third stage as a sorted list of names by their declared location. The sorted names are the names after the fourth stage. Notice that it includes dependencies over (implicitily declared) type invariants for \verb'R' and \verb'T'. Finally, the organised names are those with both name-spaces consolidated containing only names within the originally declared set. 

Furthermore, notice that the order between functions \verb'head' and \verb'tail' and type \verb'T' is ``unstable'' (\ie~they are incomparable in this case). That is, whatever order they are declared is valid, so long as they are declared before \verb'S', which has to be declared before \verb'Rec'. It is these kinds of incomparibility between certain names that makes sorting modules, where all names are declared before use can give irrelevant sorts. We try to minimise this by sorting the starting points by declaration order, but that cannot entirely prevent irrelevant sorts.  

The algorithm satisfies the invariant that the original and organised name sets are equal, and that the sorted names list contains all organised names. This way, all names are accounted for and no implicitly created names (\eg~\verb'inv_T') are part of the result. 

The translation has optional flags to generate VDM-source location information, as well as the original VDM-source, as Isabelle comments. This way, any model rearrangement is immaterial to the translation. VDM users ought not to change the Isabelle translation, and instead work on the proof obligation proof scripts.   

\begin{figure}[htbp]
    \centering
        \includegraphics[width=\textwidth,scale=0.4]{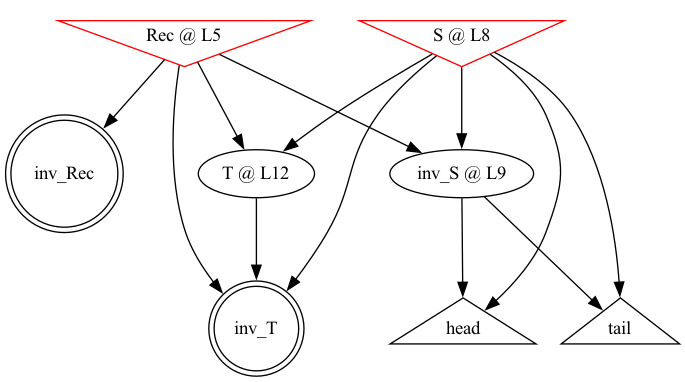}
    \caption{\texttt{M.vdmsl} definition dependencies.}\label{fig:Mdot}
 \end{figure}

\texttt{exu} also produces a \texttt{dot} file view of the dependencies, which is similar to VDMJ's order plugin output for module name dependencies (see Figure~\ref{fig:Mdot}). It depicts the user declared dependencies before sorting. It includes corresponding source line locations, where starting nodes are tagged red with an inverted triangle shape, implicitly created invariants are double circles, and (terminal) nodes without outgoing links are shaped as a triangle.  
For complex (and large) modules, such visualisation ``clues'' can be invaluable to figure out what tweaks to make in case unstable sorts produce results that are not entirely declaration before use and might generate Isabelle errors. 

\section{VDMJ and VDM-VSCode integration}\label{sec:integration}

Our plugin infrastructure follows the UNIX principle of strict separation between command line operations and their corresponding GUI outputs. The plugin architecture starts with a VDMJ \texttt{CommandPlugin} extension. This enables VDMJ console execution of the \texttt{exu} plugin via the command line. This plugin is setup, such that it will be used by all interfaces. 

For the VDMJ console interface, it follows the VDMJ convention that plugin classes are named according to the plugin call. This exposes functionality to the VDMJ debugging console like a call to a C/Java program \texttt{main} function with user-provided arguments.   

For the VDM-VSCode interface, it follows the VDM-VSCode Language Server Protocol (LSP) convention~\cite{AdvancedVSCodePaper} that the plugin class encapsulates both GUI-editor reactive functionality as well as the Debug Protocol (DAP) command line debugging. For the DAP VDM debugging session, it offers the same VDMJ console plugin commands, where any issues or errors are reported to the DAP output console itself, as opposed to any GUI visualisations of such errors. This is mostly for debugging and low-level application purposes only. Users are encouraged to go via the GUI visualisation process instead. 

For the GUI visualisation process, the \texttt{ISA} LSP plugin reacts to two LSP events: \texttt{CompleteCheckEvent} and \texttt{UnknownTranslateEvent}. The former occurs after VDMJ has typechecked all loaded VDM-SL modules within a VDM-VSCode project, whereas the latter occurs when the user explicitly chooses the project pop-up menu associated with VDM as {\verb'VDM->Translate to Isabelle'}. 

For the \texttt{CompleteCheckEvent}, the plugin performs a staggered execution of aspects of the various plugins, including execution of \texttt{exu} for checking and sorting, \textit{vdm2isa} for translation, and \textit{isapog} for proof obligation translations. The result is the output of module definition dependency graphs, translation of both VDM source and proof obligations to Isabelle, as well as any errors and warnings that might have occurred during this process. 

For the \texttt{UnknownTranslateEvent}, VDM-VSCode-wide translation options are taken into account and the same process of processing modules for translation is performed. This is somewhat wasteful in the sense that translation happens both at right after typechecking time, and when the user explicit requests are made. This is a compromise on speed and usability:~the analysis for a successful translation is tightly coupled with what the translated output is supposed to be. Thankfully the ``lag'' time is reasonable on average and can be avoided by the user (\ie~translation only occurs if the user asks via the \texttt{UnknownTranslateEvent} pop-up menu click). 

The trade off for this use case is that errors are handled in tandem and only after a translation request has happened. If the \texttt{CheckCompleteEvent} processing is fast for the loaded project, then that is a preferred route for translation, as it will keep the user informed as the model is developed. Otherwise, translation can take place at the user's request and any remaining issues can be reported and dealt with accordingly.  

For the VDM-VSCode integration, we have tried to keep all \(3\) use cases (\ie~VDMJ console, VDM-VSCode DAP console, and VDM-VSCode LSP GUI-reactivity) as independent as possible, as well as reusable as possible. This is important so that each functionality does not need to be re-encoded in slightly different ways. This has been achieved through the use of he UNIX principle of argument-based command-line execution of the plugin as a the bottom-line for all kinds of interfaces. Concretely, this means the \texttt{exu} tool is integrated within both command line and VSCode as stand-alone, independently of the Isabelle translator. 

In practice, this means each plugin responds to specific commands. This is similar to other UNIX commands (\eg~\texttt{git} has commands like \texttt{push}, \texttt{pull}, \etc). Various default options are in place, and can be changed by a properties file, command line \texttt{set} command, or VSCode properties. The property read and set process is close to VDMJ's own properties file. For VDM-VSCode, the GUI also exposes these properties through its settings interface. Ultimately, the VDM-VSCode plugin makes specific command calls to specific plugins, depending on whether it is part of the LSP's reactive behaviour  (\eg~setting yellow/red squiggles on code locations to show warnings/errors), or part of the DAP's interactive behaviour (\eg~calling specific DAP plugins on the debug console). Correspondingly, the same functionality is available through the VDMJ console via the command line. Output of translation is saved to \texttt{./.generated/isabelle} by default, though users can set it to any directory of their choice. 

\section{Results and Limitations}\label{sec:Examples}

With the module definition sorting, we have overcome the major impediment for VDM users to using the VDM to Isabelle/HOL translator. Nevertheless, there are still both VDM language restrictions of use that must be observed, such as the way certain union types of value declarations are made. Moreover, some VDM constructs are yet to be supported, such as multiple record patterns within the parameters of various declarations (\eg~declaration of invariants of explicit function signatures). This is because of the need for local contexts for record translations. We envisage that some of these restrictions will be available in future. 

Moreover, even though there is a translation strategy for various VDM statements~\cite{NimFull}, support for them is quite limited. This is mostly because their implicit specifications (\eg~the implicit updates to state as statements execute) are quite complex~\cite{10.5555/3204179.3204224}. Thus, support for VDM state and operations is quite limited. State is translated as just a special kind of record, operation signatures are available, as well as some statements (\eg~if-statements, assignments, state designators, \etc). Many other statements are missing.      

We have worked through a large subset of the VDM-SL examples, and over \(100\) test modules involving various (sometimes exotic) VDM constructs. The complete set of examples can be found at the \texttt{VDM\_Toolkit} github page\footnote{\url{https://github.com/leouk/VDM_Toolkit/tree/main/plugins/vdm2isa/src/test/resources}}.

Such excursions through the VDM-SL AST have helped to find corner cases missing from the VDM tools. For example, multiple type binds within map comprehension expressions like:
\begin{vdmsl}[frame=none,basicstyle=\ttfamily\scriptsize]
    values 
        v = { 1|-> 5 | x in set {<A>,<B>,<C>}, x in set {1,2,3} & x = <A> }
\end{vdmsl}
The type for \(x\) here ended up being the set of the union of each set's type. This is now a type error:~it should not be possible to have such implicit type union within VDM comprehensions. 

\section{Results and discussion}\label{sec:Results}

In this paper, we presented a solution to the main problem of translating VDM-SL models to Isabelle/HOL:~having to sort declaration used before they are provided. This involved various basic VDM(J) tool extensions, and  translator-specific adaptations. 

The outcome has also been integrated as part of VDM-VSCode extension~\cite{AdvancedVSCodePaper}. This should enable most users acces to the VDM to Isabelle/HOL translator without the hassle of reorganising VDM files. This is particularly important for legacy (and larger) models, where such considerations were not taken into account during development. 

We strived to keep the ecosystem stable and environment independent. As such, the same solution works in all three specific VDM environments preferred by users. Namely, the VDMJ (UNIX) command line, VDM-VSCode editor, and VDM-VSCode debuggin session. All three interfaces demonstrates the same behaviour using the same code, hence removing any kind of code repetition, thanks to the UNIX principle of command-line consistency. 

\paragraph*{Future work.}~
We plan a tighter integration between the VDMJ console plugins and VDM-VSCode GUI through the use of Code Lenses. These are user GUI-actions that trigger specific plugin events. For instance, \texttt{jUnit} testing creates a green/red circle to the left of each Java method defined as a test case (\ie~annotated with the \texttt{@Test} mark). Users can click on that to either run, debug or go to specific tests. We envisage similar code lenses for translation and different proof script strategies for VDM modules. Such strategies can be automatically tried in tandem in order to ``discover'' proofs for various VDM POs much like Isabelle's own \texttt{sledgehammer} tool.      

\paragraph*{Acknowledgements.}~
We would like to thank discussions with Peter Gorm Larsen and various of his students on different aspects (and impact of limitations) of the translator and for his encouragements on pushing various parts of tool to the surface to end users.

\bibliographystyle{splncs03}
\bibliography{order.bib}


\end{document}